%% file: arch.tex
\documentclass[conference]{IEEEtran}
\usepackage{xspace}
\usepackage{array}
\usepackage{rotating}
\usepackage{amssymb}
\usepackage{tabularx}
\usepackage{soul}
\usepackage{fancyvrb} 
\usepackage{listings}
\usepackage{booktabs}
\usepackage{multirow}
\usepackage{xurl}
\usepackage{booktabs}
\usepackage{amsmath,amsfonts,chemarrow, subcaption, url, colortbl}
\usepackage{algorithm}
\usepackage{algpseudocode}
\usepackage{tikz}
\usepackage[normalem]{ulem}
\usepackage{makecell}
\usepackage{enumitem}
\usepackage{pifont}
\usepackage{todonotes}
\usepackage[most]{tcolorbox}
\usepackage[table]{xcolor}

\definecolor{orange}{HTML}{F36E24}
\definecolor{green}{HTML}{038E79}
\definecolor{gray}{HTML}{8196A6}
\definecolor{myred}{HTML}{CE1F3A}
\definecolor{myblue}{HTML}{00619E}
\definecolor{navy}{HTML}{335E6F}

\definecolor{ExecutionColor}{HTML}{D96B00}
\definecolor{IdentityColor}{HTML}{438D4A}
\definecolor{CommunicationColor}{HTML}{3373C4}
\definecolor{ObservabilityColor}{HTML}{B57B00}
\definecolor{PersistenceColor}{HTML}{8A5AA6}

\newcommand{\BfPara}[1]{{\noindent\bfseries #1.}\xspace}

\newtcolorbox{takeawaybox}{
  colback=gray!12,
  colframe=gray!60,
  boxrule=0.5pt,
  arc=2mm,
  left=6pt,
  right=6pt,
  top=6pt,
  bottom=6pt,
  before skip=8pt,
  after skip=10pt,
  fontupper=\small
}

\definecolor{orange}{HTML}{F36E24}
\definecolor{green}{HTML}{038E79}
\definecolor{gray}{HTML}{8196A6}
\definecolor{myred}{HTML}{CE1F3A}
\definecolor{myblue}{HTML}{00619E}
\definecolor{navy}{HTML}{335E6F}

\ifCLASSINFOpdf

\else

\fi

\begin{document}

\title{Trust Without Boundaries: An Architectural Analysis of Satellite Flight Software}

\author{
\IEEEauthorblockN{
Jack Vanlyssel\IEEEauthorrefmark{1},
Gruia-Catalin Roman\IEEEauthorrefmark{1},
Kendra Cook\IEEEauthorrefmark{3},
Sazzadur Rahaman\IEEEauthorrefmark{2},
Afsah Anwar\IEEEauthorrefmark{1}
}

\IEEEauthorblockA{
\IEEEauthorrefmark{1}Department of Computer Science, University of New Mexico, Albuquerque, NM, USA
}

\IEEEauthorblockA{
\IEEEauthorrefmark{2}Department of Computer Science, University of Arizona, Tucson, AZ, USA
}

\IEEEauthorblockA{
\IEEEauthorrefmark{3}Science Applications International Corporation (SAIC)
}

}

\maketitle

\input{sections/abstract}

\input{sections/introduction}
\input{sections/background}
\input{sections/organization}

\input{sections/architecture-analysis}

\input{sections/exploiting-architectural-trust}
\input{sections/generalization}
\input{sections/discussion}

\input{sections/conclusion}

\bibliographystyle{IEEEtran}
\bibliography{references}

\end{document}

%% file: sections/abstract.tex
\begin{abstract}

As spacecraft become more software-driven and interconnected, onboard flight software is an increasingly important security boundary.
Popular flight software architectures often treat onboard components as trusted peers, simplifying integration while limiting internal isolation and access control.
We analyze NASA’s Core Flight System (cFS) to examine how authority, identity, communication, observability, and persistence are distributed across onboard components.
Using NASA’s flight-representative NOS3 simulator, we validate these weaknesses through five experiments implemented with a malicious onboard component that abuses legitimate architectural privileges.
We then compare cFS with other modular flight software frameworks to identify recurring trust assumptions and architectural weaknesses.
Our results show that a single compromised component can exploit broadly shared authority in ways that are difficult to distinguish from legitimate behavior.
We conclude with architectural implications and discuss mechanisms for strengthening internal trust boundaries in future flight software systems.

\end{abstract}

%% file: sections/introduction.tex
\section{Introduction}
\label{sec:intro}

The space industry is projected to exceed one trillion dollars by 2032~\cite{spacefoundation2025q2}. Satellites have become essential infrastructure supporting telecommunications, Earth observation, scientific experimentation, navigation, and defense~\cite{pultarova2025starlink,Tieby2024LEOCyber,SiddiqueSmallSatRevolution,Kopacz2020,nanoavionics2025smallsats,kogut2024smallsats,Sweeting2018,SONG2024104932}. To meet growing demands for rapid deployment, modern spacecraft increasingly rely on modular, component-based flight software frameworks. In this context, a component is a modular software unit that encapsulates a particular capability and interacts with the rest of the system through defined interfaces~\cite{omgUML251}. These frameworks decompose spacecraft functionality into separately developed components that implement spacecraft functions such as navigation, telemetry processing, or hardware control. These components then communicate through shared messaging infrastructure and system services. This architectural model improves maintainability and flexibility while enabling increasingly complex software stacks to be assembled from reusable components.

We argue that shared software services, while improving development efficiency, also create implicit trust relationships among onboard applications through common messaging infrastructure, system services, and runtime environments. Security therefore depends not only on preventing initial component compromise, but also on constraining the authority of components in case preventative measures fail. Yet existing satellite security research has largely focused on external attacks, communication vulnerabilities, ground-segment compromise, and implementation flaws~\cite{oligeri2020gnss,Salkield2023,Planta2024,giuliari2021icarus,Yoon2024,LiuDarkSide,viasat2022ka,peters2014noaa,donchev2024ransomware}, paying comparatively little attention to whether flight software architectures themselves enforce security boundaries. This gap is important because a compromised component may abuse legitimately granted authority without exploiting an additional vulnerability, potentially evading code review and defenses focused on initial compromise or implementation-level flaws.

To address this gap, we analyze the architecture of NASA's core Flight System (cFS), the most prolific open-source flight software framework \cite{nasa_cfs_2026}, as a representative example of modern component-based flight software. Because no single architectural mechanism determines whether a component is meaningfully constrained, we examine cFS through five complementary, security-focused "views": execution, identity, communication, observability, and persistence. Together, these views capture how authority is granted to a component, how it is exercised across the system, whether its actions can be attributed, and whether its effects survive recovery resets. This allows us to examine how the architecture as a whole establishes, propagates, and limits capabilities among onboard components.

To evaluate the practical consequences of the architectural weaknesses we identify, we implement a malicious onboard component in NASA's Operational Simulator for Space Systems (NOS3). We use this implementation to test whether these weaknesses can be exercised in a flight-representative mission environment. After confirming the weaknesses, we apply a targeted analysis to other open-source flight software frameworks to determine whether similar trust assumptions and architectural weaknesses extend beyond cFS.

Our results show that cFS grants onboard components broad authority through shared execution, unrestricted inter-component communication, and limited identity- and policy-based enforcement. As a result, the architecture provides few effective mechanisms for containing a compromised component once execution begins. Our comparative analysis further indicates that these weaknesses are not unique to cFS: other open-source flight software frameworks exhibit similar reliance on shared services, trusted components, and weak internal isolation. Together, these findings suggest that the absence of enforceable trust boundaries is a broader architectural issue in modern modular flight software that must be addressed.

\begin{figure}[t]
    \centering
    \includegraphics[width=0.9\linewidth]{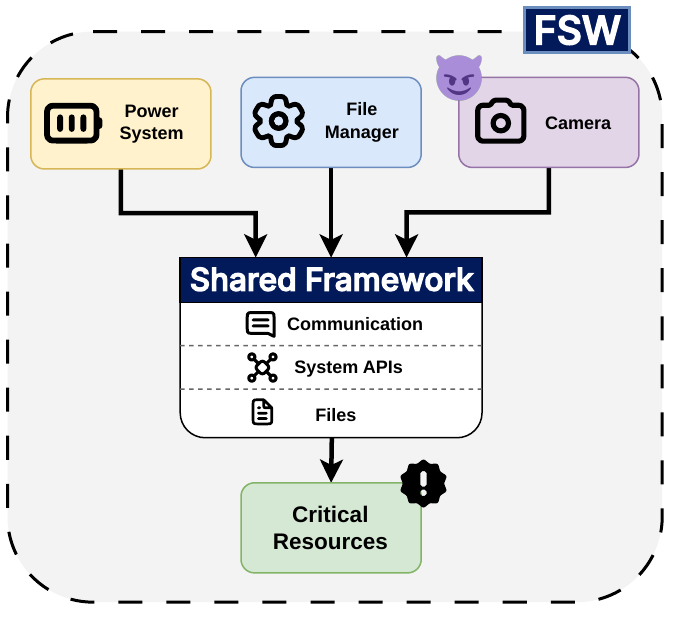}
    \caption{Architectural trust relationships in modular flight software can allow low-privilege components to influence security-critical services beyond their intended responsibilities.}
    \label{fig:flight_software}
\end{figure}

\textbf{Contributions.}
This paper makes the following contributions:

\begin{itemize}[leftmargin=*,itemsep=-1.5pt,topsep=0pt]

\item \textit{A systematic analysis of architectural weaknesses in cFS.}
We develop and apply a framework for analyzing trust in component-based flight software across execution, identity, communication, observability, and persistence. Using this framework, we identify where cFS fails to establish enforceable trust boundaries and characterize the security

consequences of those weaknesses.

\item \textit{An experimental evaluation of architectural trust assumptions.}
We experimentally validate the identified architectural weaknesses in NASA's NOS3 simulator and demonstrate system-wide influence using only architecturally permitted mechanisms.

\item \textit{A comparative analysis of open-source flight software.}
We apply our framework to additional open-source flight software architectures to determine whether the trust assumptions and enforcement weaknesses identified in cFS recur across contemporary component-based frameworks.

\item \textit{Architectural design implications.}
Based on our findings, we discuss mechanisms needed to establish enforceable trust boundaries in future flight software systems, including identity propagation, policy enforcement, and mandatory logging.

\end{itemize}

%% file: sections/background.tex
\section{Background \& Related Work} 
\label{sec:back}

\subsection{Flight Software Background}

Flight software has traditionally been designed around strict requirements for reliability, deterministic execution, resource efficiency, and fault tolerance. Frameworks such as cFS support these goals through shared services for communication, timing, event reporting, data management, and hardware access, reducing duplication and simplifying mission development. These architectures typically assume that onboard applications cooperate as trusted parts of a single software stack rather than operate as mutually untrusted elements. As a result, mechanisms used in general-purpose systems to enforce strong internal security boundaries may be absent, optional, or outside the framework itself, allowing a compromised component to exercise substantial authority through otherwise legitimate interfaces.

Table~\ref{tab:opensource_fsw} summarizes the most prominent publicly documented, open-source flight software considered in this work. Despite differences in implementation, these frameworks share recurring architectural features, including component abstractions, standardized inter-component interfaces, a shared communication infrastructure, and support for mission-specific and third-party software~\cite{Miranda2019FSWSurvey}. These features are security-relevant because they determine how components communicate, access shared services, exercise authority, and interact with mission-critical resources, making the underlying architecture central to evaluating isolation, privilege boundaries, observability, and fault containment.

\input{tables/flight-software}

\input{tables/related_work}

\subsection{Related Work}
Satellite security research has largely focused on external communication links and ground system compromise ~\cite{kang2024survey,Pietro_comm_sec,Jero_securing_stack}, typically treating the flight software stack as a trusted environment. Consequently, the security implications of trust abuse have received limited attention. To make this gap explicit, we review prior work and organize it based on the attack surface it examines.

\BfPara{External Communication and RF-Side Attacks}
The largest body of prior work models adversaries operating outside the spacecraft, targeting radio-frequency links and protocol-level communication surfaces. This includes spoofing and jamming attacks~\cite{oligeri2020gnss, Salkield2023, Planta2024}, broadcast storms and large scale denial-of-service against satellite constellations~\cite{giuliari2021icarus, Yoon2024, LiuDarkSide}, and replay and injection attacks~\cite{pavur2020tale, Yuvraj2025, LiuMindLocation}. These works largely ignore the flight software stack and assume insecurity stems from communication protocol exploits.

\BfPara{Ground Segment and Infrastructure Compromise}
A second class of work examines adversaries who pivot through terrestrial infrastructure, including ground stations, mission control software, and satellite network service providers. Representative incidents include broadband network compromise and mission control abuse~\cite{viasat2022ka,peters2014noaa} and user-terminal exploitation~\cite{smailes2023dishingdosdisablesecure}. These works also largely ignore the flight software stack and assume insecurity stems from compromising ground-side control infrastructure.

\BfPara{Onboard Software and Flight-Software Security}
Recent work has demonstrated how implementation flaws in onboard software can lead to system compromise. Donchev et al.~\cite{donchev2024ransomware} exploit a buffer overflow in a custom flight application to execute shellcode on the underlying operating system. Hansen et al.~\cite{GuardingGalaxy} demonstrate ransomware by modifying a file-management component, while Willbold et al.~\cite{Willbold2023} show that insecure telecommand interfaces, memory vulnerabilities, and unsafe update mechanisms can enable arbitrary code execution. These studies demonstrate the consequences of conventional software vulnerabilities but do not examine the broader trust relationships established by the flight-software architecture.

Other work examines specific security-relevant mechanisms within flight-software architectures. Boehm et al.~\cite{boehm2024fprime} identify weaknesses in F' external interfaces and recommend command authentication, link encryption, and secure defaults. Schalk et al.~\cite{SoftwareBus_Vulnerabilities,Vulnerabilities_SoftwareBus} demonstrate attacks against the cFS Software Communication Bus and propose bus-level detection and mitigation. Furgala et al.~\cite{FurgalaPortingNC} port cFS to seL4 to strengthen operating-system isolation, while McAmis et al.~\cite{McAmis2026CompromisedPeripheral} show that compromised peripherals can manipulate data and timing in cFS-based systems. These efforts strengthen or evaluate individual interfaces and mechanisms, but do not analyze how trust and authority propagate across the flight-software architecture as a whole.

\textbf{Research Gap.}
Taken together, these works identify vulnerabilities across flight software, including weaknesses in the Software Bus, operating system, peripheral interfaces, and external communication paths. As summarized in Table~\ref{tab:related_work}, however, prior studies primarily evaluate individual attack surfaces and defensive mechanisms in isolation. Consequently, it remains unclear whether these vulnerabilities stem from implementation flaws or from architectural trust assumptions that span the flight software stack. This work addresses that gap by evaluating the trust relationships that determine the authority available to an onboard component and the extent of its influence over the rest of the system.






%% file: tables/flight-software.tex
\begin{table}[H]
\centering
\caption{Prominent Open-source Flight Software available today \cite{Miranda2019FSWSurvey}.}
\label{tab:opensource_fsw}
\small
\scalebox{0.9}{
\begin{tabular}{lll}
\hline
\textbf{Platform} & \textbf{Org.} & \textbf{Mission Adoption} \\
\hline

\colorbox{IdentityColor}{\textcolor{white}{cFS}}
  & NASA
  & 40+ NASA Missions \\

\colorbox{gray}{\textcolor{white}{F$^\prime$}}
  & NASA 
  & Mars Helicopter, Mars Rover, CubeSats \\

\colorbox{gray}{\textcolor{white}{NanoSat MO}}
  & ESA
  & OPS-SAT, PhiSat \\

  \colorbox{gray}{\textcolor{white}{COrDeT-C2}}
  & ESA
  & CHEOPS Telescope \\

  \colorbox{gray}{\textcolor{white}{KubOS}}
  & Kubos
  & limited public flight evidence \\

\hline
\end{tabular}}
\end{table}

%% file: tables/related_work.tex
\begin{table*}[t]
\centering
\small
\setlength{\tabcolsep}{6pt}
\renewcommand{\arraystretch}{1.15}
\begin{tabularx}{\textwidth}{
    >{\raggedright\arraybackslash}p{3.2cm}
    >{\raggedright\arraybackslash}X
    >{\raggedright\arraybackslash}X
    >{\raggedright\arraybackslash}p{4.0cm}}
\toprule
\textbf{Research Area} &
\textbf{Representative Focus} &
\textbf{Limitation Relative to This Work} &
\textbf{Related Works} \\
\midrule

Comms and RF-side attacks &
RF links, spoofing, jamming, replay &
Assumes trusted onboard software. &
\cite{oligeri2020gnss,giuliari2021icarus,pavur2020tale,LiuMindLocation} \\

Ground-segment compromise &
Ground stations, network infrastructure &
Focuses on offboard compromise. &
\cite{viasat2022ka,peters2014noaa,smailes2023dishingdosdisablesecure} \\

Software vulnerabilities &
Memory corruption, malicious apps &
Targets implementation flaws. &
\cite{donchev2024ransomware,GuardingGalaxy,Willbold2023} \\

Software security mechanisms &
OS isolation, interface and message security &
Examines individual mechanisms. &
\cite{boehm2024fprime,SoftwareBus_Vulnerabilities,Vulnerabilities_SoftwareBus,FurgalaPortingNC,McAmis2026CompromisedPeripheral} \\

\midrule

\rowcolor{IdentityColor!20}
\textbf{This work} &
\textbf{Architectural trust relationships} &
---&
--- \\

\bottomrule
\end{tabularx}
\caption{Relationship between prior flight-software security research and this work.}
\label{tab:related_work}
\end{table*}

%% file: sections/organization.tex
\section{Approach and Threat Model}
\label{sec:approach}

\begin{figure*}[t]
    \centering
    \includegraphics[width=0.78\linewidth]{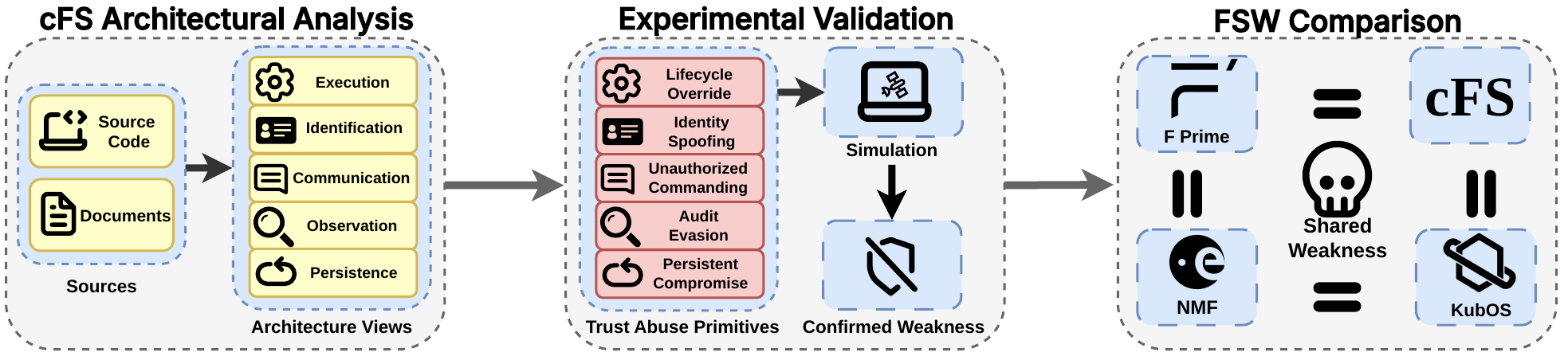}
    \caption{We organize our efforts into three steps: First, we perform a systematic analysis of flight software architectures. Next, we develop proof-of-concept attacks to understand the violations of established security expectations. Finally, we measure the generalizability of our findings to other flight software frameworks.} 
    
    \label{fig:organization}
\end{figure*}

To determine whether modern flight software architectures provide enforceable trust boundaries between onboard software components, we combine architectural analysis, experimental validation, and comparative analysis, as summarized in Figure~\ref{fig:organization}.

\subsection{High-level Approach}

\textbf{Objective 1: Characterize Architectural Trust.}
We systematically decompose NASA's core Flight System (cFS) into its principal architectural elements of execution, identity, communication, observability, and persistence to determine whether the architecture enforces least privilege and isolation between onboard components. This analysis is derived from publicly available documentation, source code, startup configurations, and reference mission deployments~\cite{cfe_app_dev_guide,cfe_users_guide,nasa_cfs_etd,fprime2025nasa,nos3,fpmission_fs_picamera,nasa_arducam_cfs,nos3_cpu1_cfe_es_startup,nos3_configs}.

\textbf{Objective 2: Validate Architectural Behavior.}
We experimentally evaluate whether a compromised cFS application can use legitimate framework mechanisms to produce security-relevant effects outside its intended role in NASA's NOS3 flight-representative environment.

\textbf{Objective 3: Assess Generality.}
Finally, we examine other open-source flight software using our framework to determine whether the architectural trust assumptions identified in cFS recur across contemporary flight software or are an implementation issue in a single framework.

\subsection{Threat Model}

We assume that an adversary has introduced malicious code into one component of an otherwise correctly functioning flight software system. The code may originate from a compromised dependency, malicious update, insider action, or another compromise mechanism. Because this study focuses on the architectural consequences of a compromised component, the initial compromise mechanism is outside its scope.

\noindent\textbf{Attacker objective.}
The attacker's general objective is to use the compromised component's legitimate access to affect resources, applications, or operations outside its intended role. The specific objective and success condition evaluated by each attack are defined in Section~\ref{sec:experimental_validation}.

\noindent\textbf{Attacker capabilities.}
The attacker can write malicious code that executes within the legitimately assigned context of a component. The attacker may invoke services, access resources, and use communication mechanisms ordinarily available to that component. The attacker does not exploit additional implementation vulnerabilities, modify the operating system or flight software framework, bypass enforced controls, or obtain privileges beyond those assigned to the compromised component.

Under this model, we examine whether the flight software architecture meaningfully constrains a compromised component after it has begun executing.

%% file: sections/architecture-analysis.tex
\section{Architecture Analysis}
\label{sec:architecture analysis}

\BfPara{Architectural Analysis Methodology}

For a software architecture to provide a meaningful security boundary, it must limit what each component can do, identify which component performed an action, control how components interact with the rest of the system, record enough information to investigate security incidents, and ensure that unauthorized changes do not survive recovery. These requirements follow established principles of least privilege, mediated access, accountability, and trusted recovery~\cite{1451869,nist80053r5,ross2021cyberresilient}.

Based on these requirements, we organize our analysis into five security-focused "architectural views" ~\cite{clements2005views}. Each view addresses a distinct question about whether the flight software establishes and enforces boundaries around an executing component:

\begin{enumerate}
\item \textbf{Execution view:} Where does a component run, and what privileges and resources does it receive?
\item \textbf{Identity view:} Does the system assign a distinct identity to each component and preserve that identity across interactions?
\item \textbf{Communication view:} How can components interact with other components and shared services, and are those interactions subject to enforceable policy checks?
\item \textbf{Observability view:} Are component actions recorded in a way that allows them to be detected and attributed?
\item \textbf{Persistence view:} Which component-controlled effects survive restart, and does recovery return the system to a trusted state?
\end{enumerate}

For each architectural view, we first review framework documentation and API references to identify the major components, interfaces, and intended security behavior. We then inspect the corresponding source code, startup scripts, and reference mission configurations to determine how those mechanisms are implemented, what capabilities are available to onboard applications, and whether restrictions are enforced in practice or depend on trusted behavior or mission-specific configuration~\cite{nasa_cfs_github,fprime2025nasa,nasa_cfs_etd,hall2023nasa_cfs,werner2025cfsupdate,nos3_configs}. We organize the resulting findings according to the five architectural views and corroborate them across documentation, implementation, and configuration evidence. Findings central to our conclusions are subsequently tested in Section \ref{sec:experimental_validation}.

\subsection{\textbf{Execution View}}
\label{sec:execution view}
In cFS, components are typically implemented as a flight application (e.g., camera\_app.c) together with a collection of supporting source files and, when required, a hardware device driver. Applications serve as the primary units of execution and resource allocation. The execution view therefore determines how these applications are created, scheduled, and granted access to system resources. The Core Flight Executive (cFE) acts as the central runtime environment, loading and coordinating independently developed applications.

The cFE provides five core services: Executive Services (ES) manages application lifecycle and execution state; the Software Bus (SB) provides inter-application messaging; Event Services (EVS) records ground-visible application events; Table Services (TBL) manages dynamically loadable configuration data; and Time Services (TIME) maintains and distributes spacecraft time.

To determine what authority an onboard component receives, we examine its execution environment. Specifically, we analyze how cFE creates applications, how the underlying operating system runs and isolates them, and which system services and resources they can access.

\textit{(1) Instantiation:} Applications are instantiated at boot by ES through startup scripts that specify which binaries to load and execute. ES dynamically loads each listed binary and creates a corresponding OS thread. This process performs no cryptographic verification, provenance checking, or privilege assignment. Any referenced binary is therefore automatically executed as a trusted system component.

\textit{(2) Execution:} Once instantiated, all applications, including core services, execute as OS-level tasks within a single shared address space. cFS provides no process isolation, memory protection, or privilege separation between components. Startup entries provide no mechanism for enforcing execution boundaries. As a result, core services and the components they manage operate under identical runtime conditions.

\textit{(3) Authority:} Applications interact with cFE through the standard \texttt{cfe.h} interface, which provides access to messaging, command handling, event reporting, and time synchronization and is required for normal operation. This interface also exposes privileged lifecycle, configuration, and system-management functions, enabling applications to invoke cFE services ranging from routine messaging to terminating or restarting other components, as illustrated in Table~\ref{tab:cfe_apis}. As a result, all applications are granted broad, system-wide authority over both system resources and application lifecycles, with no architectural mechanism for enforcing least privilege or hierarchical control.

\begin{table}[h]
\centering
\scalebox{0.85}{
\begin{tabular}{lll}
\textbf{Service} & \textbf{Representative API} & \textbf{Architectural Authority} \\
\hline
ES   & \texttt{CFE\_ES\_RestartApp()}  & Restart applications \\
SB   & \texttt{CFE\_SB\_TransmitMsg()} & Transmit internal packets \\
EVS  & \texttt{CFE\_EVS\_SendEvent()}  & Down-link event string \\
TBL  & \texttt{CFE\_TBL\_GetAddress()}   & Obtain pointer to config \\
TIME & \texttt{CFE\_TIME\_SetTime()}   & Modify system time \\
\end{tabular}}
\caption{Powerful cFE API calls accessible to all apps.}
\label{tab:cfe_apis}
\end{table}

Together, these mechanisms place all core services and mission applications within a single execution domain. Applications are loaded without integrity or provenance checks, execute without isolation, and inherit broad system authority through mandatory interfaces. As a result, cFS provides no meaningful internal trust boundaries among components. Once instantiated, each application operates as peer tasks within a shared address space and receives broad access to system services and resources, leaving no meaningful mechanism for enforcing least privilege between components.


\subsection{\textbf{Identity View}}
\label{sec:identification view}

Least privilege enforcement requires that system actions be attributable to executing entities. Thus, we examine how cFS assigns and maintains execution identity.

\textit{Identity Within cFE:} At startup, ES initializes a global registry that binds each application and its corresponding OS thread to a unique Application Identifier (AppID). Each application is associated with exactly one AppID, which serves as the unit of lifecycle management and control within ES. This binding is accessible through ES interfaces and remains fixed for the lifetime of the application instance. As a result, cFE maintains a centralized and consistent internal identity.

\textit{Identity Outside cFE:} AppIDs are not propagated beyond cFE. As a result, inter-application actions are decoupled from execution identity, allowing any application to issue commands, publish telemetry, or access shared services without identity-based restriction. Although cFS assigns each application a distinct AppID within cFE, that identity is not preserved across messaging and shared-service interactions, preventing reliable authentication, authorization, and attribution.


\subsection{\textbf{Communication View}}
\label{sec:communication view}

System coordination depends on the exchange of commands and telemetry across components and trust boundaries. The communication view therefore determines whether interactions are authenticated, authorized, and attributable.

\textit{Internal Messaging and Device Access:}
The Software Bus (SB) provides the primary communication substrate in cFS, routing commands and telemetry among applications based on Message Identifiers (MIDs). As observed in other works, the SB performs only minimal syntactic validation and does not authenticate message producers, authorize publications or subscriptions, or enforce trust relationships between communicating applications~\cite{cfe_app_dev_guide,SoftwareBus_Vulnerabilities, Vulnerabilities_SoftwareBus}. Consequently, any application capable of publishing a syntactically valid message can influence subscribed components or device drivers.

Because the Software Bus does not preserve sender identity, downstream services cannot verify message provenance or enforce sender-specific authorization. The Software Bus therefore places internal communication within a single trust domain: rather than limiting authority between components, it allows one component’s authority to propagate through legitimate communication paths and affect otherwise unrelated parts of the system.

\textit{External Messaging:}
Previous work has demonstrated that unsecured cFS configurations concentrate external command validation at the Command Ingest (CI) application. Once CI forwards a command onto the Software Bus, the architecture provides no general mechanism for authenticating its origin or enforcing sender-specific authorization~\cite{SoftwareBus_Vulnerabilities,Vulnerabilities_SoftwareBus}.

Although cFS has a cryptography component that can provide link-layer authentication and integrity protection, it operates only on explicitly routed traffic and does not bind onboard component identity or enforce communication policy. It also does not extend to the software bus's internal communication, only operating around the radio interface. Consequently, it can secure external transport but provides no internal protections.

These mechanisms provide no meaningful security for communications within the architecture. As a result, applications can communicate with components, services, and devices through the Software Bus without sender authentication or enforceable policy checks, allowing their authority to propagate throughout the system.




\subsection{\textbf{Observability View}}
\label{sec:observability view}

After deployment, operators have no direct visibility into onboard storage, memory, and runtime state. Detection and attribution of anomalous behavior then depend largely on downlinked telemetry and logs. cFS provides three primary logging mechanisms—system, performance, and event logs—which together determine whether component activity can be observed, attributed, and trusted.

\textit{System and Performance Logs:} ES maintains diagnostic and performance logs that record string messages and timing markers from calling applications. These logs are opt-in and capture limited diagnostic information such as when apps start and when they crash. They do not record compromise indicators such as inter-application messaging, device access, or file activity. In addition, they are not automatically downlinked to operators and can be cleared through exposed interfaces.

\textit{Event Logs:} EVS maintains an event log that applications populate using EVS functions. Unlike system and performance logs, events embed the sender's AppID, inheriting cFE's stronger identity semantics. Unlike system and performance logs, events include explicit AppID attribution and event metadata. EVS stores events locally in memory and automatically downlinks logs to operators, providing greater visibility than ES logging.

Consistent with the broad authority granted to applications in the execution model, EVS exposes commandable filtering and suppression mechanisms that allow events to be selectively discarded before logging or transmission according to rules defined by any application. EVS interfaces also permit callers to emit events under arbitrary AppIDs and erase local logs, further weakening the visibility and attribution guarantees on which operators rely.

Consequently, because logging is incomplete, application-controlled, and susceptible to suppression, erasure, and identity spoofing, cFS does not provide a mandatory audit trail that reliably detects and attributes component actions.


\subsection{\textbf{Persistence View}}
\label{sec:persistence view}

If a satellite is compromised, as a last line of defense, operators can rely on resets to restore a known-good state. The effectiveness of these measures depends on whether reinitialization can restore a trusted baseline.

cFS preserves deployment state across power cycles and processor resets. Startup scripts stored in non-volatile memory are automatically replayed at boot, causing all listed binaries to relaunch without operator validation. As shown in Section \ref{sec:execution view}, ES also exposes lifecycle management interfaces, such as \texttt{CFE\_ES\_RestartApp()}, that allow applications to control restart behaviors programmatically

In addition to executable persistence, cFS provides a Critical Data Store (CDS) that can be used to preserve application memory across resets. Applications are free to register, modify, and restore CDS blocks through ES interfaces, allowing internal application state to persist indefinitely.

These mechanisms weaken reset-based recovery because cFS may reload compromised code, configuration, or CDS state without independent verification. Consequently, a restart cannot guarantee restoration to a known-good state.


\begin{figure}[t]
    \centering
    \includegraphics[width=0.8\linewidth]{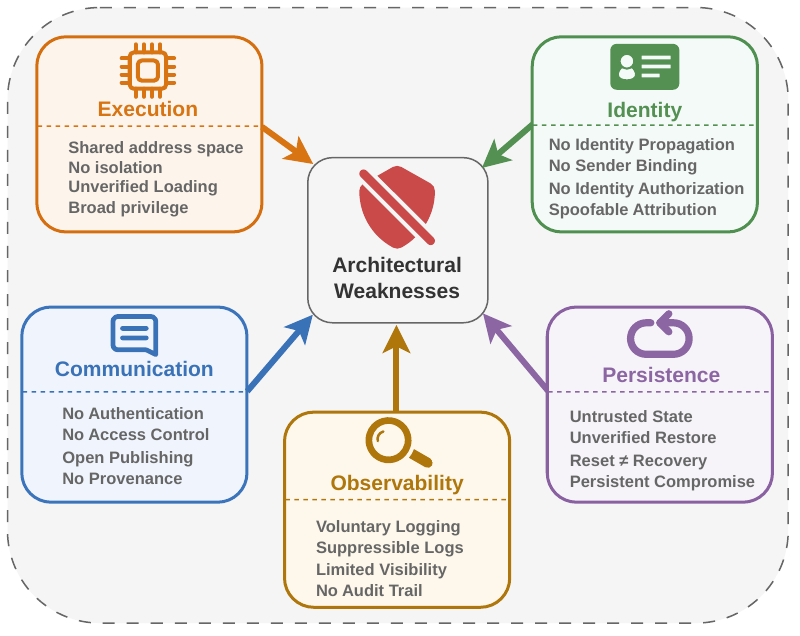}
    \caption{Architectural analysis dimensions and the recurring properties that weaken internal trust boundaries across execution, identity, communication, observability, and persistence.}
    \label{fig:arch_issues}
\end{figure}

\BfPara{Cross-View Findings}
Based on our analysis, cFS places onboard applications within a shared trust domain. Applications execute with broad authority that is difficult to attribute. As illustrated in Figure~\ref{fig:arch_issues}, these weaknesses reinforce one another: a compromised application can use legitimate interfaces to affect unrelated components, obscure its actions, and resume operation after restart. The resulting risks are therefore architectural, arising from the assumption that onboard applications will behave correctly.

%% file: sections/exploiting-architectural-trust.tex
\section{Experimental Validation of Architectural Predictions}
\label{sec:experimental_validation}

\begin{figure*}[t]
\centering
\includegraphics[width=0.7\linewidth]{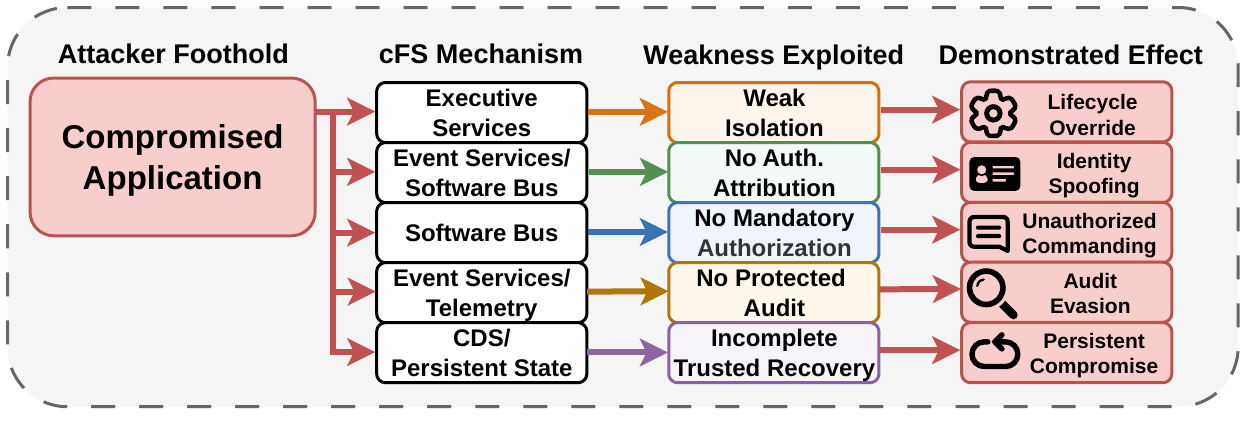}
\caption{Experimental validation pathways from attacker foothold to cFS mechanism, architectural weakness, and demonstrated effect. Each row shows how a compromised application uses a legitimate framework mechanism to produce an effect outside its intended mission role.}
\label{fig:experiments}
\end{figure*}

The preceding analysis identifies capabilities and control pathways available to onboard applications from cFS documentation, source code, and reference configurations. Static evidence alone, however, does not establish whether these capabilities remain usable in an integrated flight environment, where mission configuration, runtime checks, ownership rules, or receiver-side validation may impose additional constraints. We therefore experimentally test whether the identified weaknesses remain exploitable in practice.

\subsection{Experimental Platform}

We evaluate cFS using NASA's NOS3, a flight-representative virtual spacecraft environment publicly released through NASA's GitHub repository~\cite{nos3}. NOS3 integrates flight software and a ground station with software models of spacecraft hardware, including radios, thrusters, power systems, cameras, and navigation sensors. Although these models replace physical devices, NOS3 preserves the software architecture, application privileges, messaging interfaces, and trust relationships relevant to our analysis, allowing us to test whether identified weaknesses remain usable across component and device boundaries.

\subsection{Methodology}

Under the threat model in Section~\ref{sec:approach}, we implement a normally integrated cFS application that exercises five representative trust-abuse primitives using only documented interfaces and existing application privileges. Each primitive is command-activated for experimental control, with a defined technical objective and observable success condition. An experiment succeeds when the application produces the specified effect outside its intended role without modifying the target or framework or obtaining additional privileges.

We also assess the breadth of exposure across the 18 operational components in the default NOS3 configuration, spanning imaging, sensing, navigation, communication, security, and onboard data processing. For each component, we determine whether its normal interfaces or resources satisfy the preconditions for one or more validated primitives.
We verify execution and target effects using system logs, telemetry, command responses, application state, and deviations from normal component behavior.

\subsection{Experiments}

Figure~\ref{fig:experiments} summarizes the five experiments, each testing whether a normally integrated application can affect resources outside its intended role.

\textbf{(1) Peer Lifecycle Override.}
The attacker seeks to disrupt mission operation or disable a component that could interfere with malicious activity. Our experimental application used Executive Services to delete, restart, and reload an unrelated mission application. Telemetry, ES status, and expected output confirmed the state change. Because the application neither owned nor supervised the target and required no additional privileges, the experiment demonstrates cross-application lifecycle control through legitimate interfaces.

\textbf{(2) Sender Identity Discontinuity.}
The attacker seeks to impersonate a trusted application or conceal the source of an action. In one mode, the application transmits a command with an arbitrary message ID, function code, and payload, while the receiver receives no authenticated producer identity. In another, it uses another application's AppID to emit a telemetry event attributed to that application. Thus, packet metadata and AppIDs do not reliably identify the originating application.

\textbf{(3) Unauthorized Command Propagation.}
The attacker seeks to manipulate spacecraft or device state outside the compromised application's mission role. The experimental application published a valid command for an unrelated device through the Software Bus, which accepted the command and changed state. This confirms that authorization depended on packet format and routing rather than authenticated sender identity or application-specific permissions.

\textbf{(4) Operator-View Equivalence.}
The attacker seeks to perform an unauthorized action without revealing its originating application. We compared the same device command issued by the intended controller and by the experimental application. Both produced the same observable effect, while available EVS, ES, performance, telemetry, Software Bus, command-response, and ground-station data did not reliably distinguish their origins. The experimental application could further obscure attribution by suppressing, altering, or assigning misleading identities to event reports.

\textbf{(5) Compromise Replay After Reset.}
The attacker seeks to preserve malicious influence across a processor reset without requiring another external command. The experimental application stored attacker-controlled state, used it to alter runtime behavior, and issued a processor reset. After restart, it recovered the state and automatically resumed the altered behavior without further attacker input. This demonstrates that component-controlled persistent state can reactivate compromised behavior during normal initialization.

\subsection{Validation Results}

All five primitives produced their predicted security-relevant effects while operating as normally integrated cFS applications, confirming that the capabilities identified by static analysis remain reachable in an integrated flight configuration. Once an application was compromised, legitimate framework mechanisms were used to extend its influence beyond the intended component boundary.

Across the 18 operational components, every application was susceptible to at least one of the evaluated attacks (Table~\ref{tab:all_apps_vulnerable}). Not every primitive was applicable to every component: some applications do not control an associated device or lack the specific interface or resource required by a given attack. In these cases, the absence of an exploitable path reflects the component's functionality rather than an architectural security boundary. Where the necessary attack surface was present, the corresponding primitive succeeded. We do not claim that every component supports every primitive or that all mission configurations expose identical paths. Rather, we found that components performing diverse mission functions depend on the same shared architectural assumptions, making trust-enabled attacks possible when the corresponding preconditions are present.

\input{tables/experiments}

%% file: tables/experiments.tex
\begin{table}[h]
\centering
\caption{Systemic Architectural Trust Exposure Across Flight Components.}
\label{tab:all_apps_vulnerable}
\small
\begin{tabular}{l}
\toprule
\textbf{Components Affected by at Least One Attack Class.} \\
\midrule
ArduCam, CryptoLib, ADCS, CSS, EPS, FSS, IMU, MAG, PiCam \\
Radio, RW, ST, Thruster, Torquer, MGR, GNSS, OnAIR, SYN \\
\bottomrule
\end{tabular}
\end{table}

%% file: sections/generalization.tex
\section{Generalizability}
\label{sec:generalization}

To determine if similar architectural issues are generalizable across different modular frameworks, we examine four additional open-source flight software projects. 
For each, we assess how software components are instantiated, communicate, establish identity, and whether the architecture provides enforceable mechanisms for least privilege and isolation. 
This comparison focuses on documented architectural properties, supplemented by targeted source-code inspection, rather than a full source-level audit of each framework.

\input{tables/cross_framework}

\subsection{F Prime}
F Prime (F´) is a component-based flight software framework in which software components are connected through strongly typed communication ports defined as part of the system architecture~\cite{fprime_release,fprime_core_constructs}. These connections are typically fixed when the flight software is built. Components can communicate either by directly invoking another component or by placing requests into a queue for later execution, depending on the type of communication port~\cite{fprime_topology,fprime_core_constructs}.

\subsubsection{Architectural Trust Assessment}
\begin{itemize}[leftmargin=*]

\item \textbf{Execution:} Components may use independent threads and queues, but components within one deployment typically share the same process, memory, and operating-system privileges~\cite{fprime_core_constructs,fprime_multicore,fprime_osal}. Strong isolation requires separate deployments and externally enforced process or platform boundaries~\cite{fprime_multicore}.

\item \textbf{Identity:} F´ assigns identifiers to components for connecting them and routing messages, but these identifiers do not securely verify which component performed an action~\cite{fprime_dictionary,fprime_cmd_dispatcher,fprime_core_constructs}. When multiple components run in the same process, the system can generally identify the process responsible but cannot reliably attribute the action to a specific component~\cite{fprime_multicore,fprime_osal}.

\item \textbf{Communication:} Typed ports and topology connections restrict intended communication, while command opcodes are routed through a command dispatcher~\cite{fprime_core_constructs,fprime_cmd_dispatcher}. The reviewed framework does not provide mandatory per-call authorization, authenticated component channels, or cryptographic protection in its standard communications protocol~\cite{fprime_multicore,fprime_osal,fprime_protocol}.

\item \textbf{Observability:} F´ provides events, telemetry, command responses, health monitoring, and configurable event filtering~\cite{fprime_event_manager}. However, it does not document a mandatory tamper-resistant audit trail or unforgeable component attribution~\cite{fprime_event_manager,fprime_multicore}.

\item \textbf{F´ Persistence:} F´ reloads parameter values explicitly saved to its parameter database and checks uploaded files for corruption, but whole-system verification and restoration to a known-good software state are left to the deployment platform~\cite{fprime_prmdb,fprime_file_uplink}.

\end{itemize}

\noindent\textbf{Recurring weaknesses.}
Like cFS, F´ typically lacks strong isolation between components in the
same deployment, authenticated attribution of component actions, mandatory
authorization at service boundaries, protected auditing, and
framework-level recovery to a known-good state.

\subsection{KubOS}

KubOS is a Linux-based flight platform in which mission applications and services run as independent processes. Services are selected at build time, and applications run directly or through scheduling services. Components primarily communicate through GraphQL over HTTP, while the communications framework routes ground-originated GraphQL and UDP traffic internally~\cite{kubos_design,kubos_ecosystem,kubos_service_config,kubos_comms_framework}.

\subsubsection{Architectural Trust Assessment}
\begin{itemize}[leftmargin=*]

\item \textbf{Execution:} Applications run as separate Linux processes, but least-privilege controls and exclusive service-mediated device access are deployment-dependent~\cite{kubos_design,kubos_app_service,kubos_hardware_services,kubos_mission_development}.

\item \textbf{KubOS Identity:} KubOS assigns processes operating-system identities, including process and user IDs, but its documented service interface does not preserve that identity across requests, limiting reliable attribution and identity-based authorization of shared services~\cite{kubos_monitoring,kubos_service_config,kubos_graphql}.

\item \textbf{Communication:} Configuration, GraphQL schemas, and packet routing mediate communication, but mandatory authentication, authorization, message integrity, and replay protection are not documented~\cite{kubos_service_config,kubos_graphql,kubos_comms_framework}.

\item \textbf{Observability:} KubOS centralizes logs, monitoring, and telemetry, but does not document trusted caller attribution or tamper-evident auditing~\cite{kubos_logging,kubos_monitoring,kubos_monitor_service,kubos_telemetry_database}.

\item \textbf{KubOS Persistence:} KubOS detects boot failures and kernel corruption and can restore current, previous, or base operating-system images; however, its user-data partition survives recovery, so user-controlled files and configuration are not necessarily returned to a known-good state~\cite{kubos_recovery,kubos_upgrade,kubos_service_config}.

\end{itemize}

\noindent\textbf{Recurring weaknesses.}
KubOS provides stronger process separation than cFS, but still shares its
lack of preserved caller identity, mandatory service authorization,
tamper-resistant auditing, and complete recovery of security-relevant
persistent state.

\subsection{NanoSat MO Framework}

The NanoSat MO Framework (NMF) deploys flight functionality as separate applications, each executed in its own operating-system process. Applications access platform, monitoring, archival, discovery, and software-management services through a shared Supervisor and may also provide services to one another. The installed application set defines the deployment, while the Apps Launcher starts and manages each process~\cite{coelho2016portability,nmf_app_development,nmf_sm_readme}.

\subsubsection{Architectural Trust Assessment}
\begin{itemize}[leftmargin=*]

\item \textbf{Execution:} Applications run in separate processes, optionally under distinct users, but sandboxing and resource confinement are not mandatory~\cite{nmf_sm_readme,nmf_apps_launcher_source}.

\item \textbf{NMF Identity:} NMF can launch applications under configured operating-system users and registers each application as a named service provider, but the reviewed documentation does not establish that this identity is authenticated and preserved across service requests~\cite{nmf_apps_launcher_source,nmf_app_development,coelho2016portability}.

\item \textbf{Communication:} Applications access services through the NanoSat MO Connector and Directory, but mandatory authorization, sender authentication, and message integrity are not documented~\cite{nmf_app_development,coelho2016portability}.

\item \textbf{NMF Observability:} The NMF Supervisor associates application output with the launched application, and some service operations are stored in a central archive, providing useful operational visibility without establishing a complete, tamper-resistant record of all component actions~\cite{nmf_sm_readme,nmf_apps_launcher_provider_source,nmf_repository}.

\item \textbf{Persistence:} Application packages and archived state persist across restarts, while signature enforcement, verified boot, rollback protection, and known-good recovery are not documented~\cite{nmf_app_packaging,nmf_sm_readme,nmf_apps_launcher_source}.

\end{itemize}

\noindent\textbf{Recurring weaknesses.}
NMF provides process-level separation but, like cFS, does not establish
mandatory confinement, authenticated attribution across service requests,
authorization enforcement, protected auditing, or trusted recovery.

\subsection{CORDET-C2}

CORDET-C2 is a C-based service-oriented flight software framework in which applications exchange commands and reports through framework-defined interfaces. A system may consist of multiple CORDET applications running on the same or different computers and communicating through external middleware~\cite{cordet_c2_user_manual,cordet_c2_repository}.

\subsubsection{Architectural Trust Assessment}
\begin{itemize}[leftmargin=*]
\item \textbf{Execution:} CORDET components are scheduler-invoked passive objects. Within an individual CORDET application, components share an execution environment without documented component-level isolation~\cite[Sec.~22.1]{cordet_c2_user_manual}.

\item \textbf{Identity:} Components and packets use numeric identifiers for routing and management, but these are logical rather than authenticated identities~\cite[Secs.~6.2 and 10.1.3]{cordet_c2_user_manual}.

\item \textbf{Communication:} CORDET mediates commands and reports and performs framework-defined lifecycle and acceptance checks, while packet representation and additional integrity checks are application-defined. Sender authentication, ACLs, and cryptographic integrity are not documented as framework-level mechanisms~\cite[Secs.~10 and 16]{cordet_c2_user_manual}\cite[Secs.~9 and 16]{cordet_c2_user_manual}.

\item \textbf{Observability:} Registries and error/status reporting provide operational visibility; mission implementations may add event and FDIR mechanisms, but protected attribution and tamper-resistant auditing are not documented as framework-level mechanisms~\cite[Secs.~15, 20, and 23]{cordet_c2_user_manual}\cite{cechticky2015cheops,ottensamer2016cheops}.

\item \textbf{CORDET-C2 Persistence:} CORDET-C2 primarily maintains volatile command and report state and delegates application reset behavior to mission-specific code, so trusted restoration of executable software, configuration, and persistent mission state remains outside the framework’s guarantees~\cite{cordet_c2_user_manual}.

\end{itemize}

\noindent\textbf{Recurring weaknesses.}
CORDET-C2 closely resembles cFS in its shared execution environment,
logical rather than authenticated identities, lack of mandatory
authorization, unprotected operational records, and application-defined
recovery.

\subsection{Summary}
Across cFS, F´, KubOS, NMF, and CORDET-C2, componentization does not guarantee strong security isolation. Shared-process frameworks lack component-level boundaries, while process-based frameworks depend on deployment-specific privilege and confinement controls. Across both models, identifiers, communication checks, logging, and persistence mechanisms support operation but generally do not provide authenticated identity, mandatory authorization, trusted attribution, or guaranteed recovery to a known-good state. These controls are therefore more robust when enforced consistently by the framework or protocol rather than delegated to individual applications, consistent with Saltzer and Schroeder's principles of least privilege and complete mediation~\cite{1451869}.

%% file: tables/cross_framework.tex
\begin{table*}[t]
\centering
\caption{Recurring architectural weaknesses across the evaluated flight-software frameworks. “Yes” indicates that the weakness is present; “Partial” indicates partial or deployment-dependent mitigation.}
\label{tab:cross_framework_weaknesses}
\small
\renewcommand{\arraystretch}{1.15}

\begin{tabular}{
    @{}l
    >{\columncolor{ExecutionColor!7}}c
    >{\columncolor{IdentityColor!7}}c
    >{\columncolor{CommunicationColor!7}}c
    >{\columncolor{ObservabilityColor!7}}c
    >{\columncolor{PersistenceColor!7}}c
    @{}
}
\toprule

\textbf{Framework}
&
\cellcolor{ExecutionColor!22}
\shortstack{\textbf{Weak}\\\textbf{Isolation}}
&
\cellcolor{IdentityColor!22}
\shortstack{\textbf{No Authenticated}\\\textbf{Attribution}}
&
\cellcolor{CommunicationColor!22}
\shortstack{\textbf{No Mandatory}\\\textbf{Authorization}}
&
\cellcolor{ObservabilityColor!22}
\shortstack{\textbf{No Protected}\\\textbf{Audit}}
&
\cellcolor{PersistenceColor!22}
\shortstack{\textbf{Incomplete}\\\textbf{Trusted Recovery}}
\\

\midrule

cFS
& Yes
& Yes
& Yes
& Yes
& Yes \\

F$^\prime$
& Yes
& Yes
& Yes
& Yes
& Yes \\

KubOS
& Partial
& Yes
& Yes
& Yes
& Partial \\

NMF
& Partial
& Yes
& Yes
& Yes
& Yes \\

CORDET-C2
& Yes
& Yes
& Yes
& Yes
& Yes \\

\bottomrule
\end{tabular}
\end{table*}

%% file: sections/discussion.tex
\section{Discussion}
\label{sec:discuss}

Our analysis identifies recurring weaknesses in flight-software architectures that assume components will behave according to their intended roles. Prior work has introduced important security improvements, including authenticated communication, OS-level isolation, capability-based access control, protected logging, secure boot, and authenticated updates. These mechanisms address specific portions of the attack surface, but they do not collectively constrain the authority of a compromised onboard component across framework services, inter-component communication, hardware interfaces, and persistent state. Our findings identify where these protections remain incomplete and provide an architectural basis for extending and integrating existing defenses.

\begin{figure}[t]
    \centering
    \includegraphics[width=0.75\linewidth]{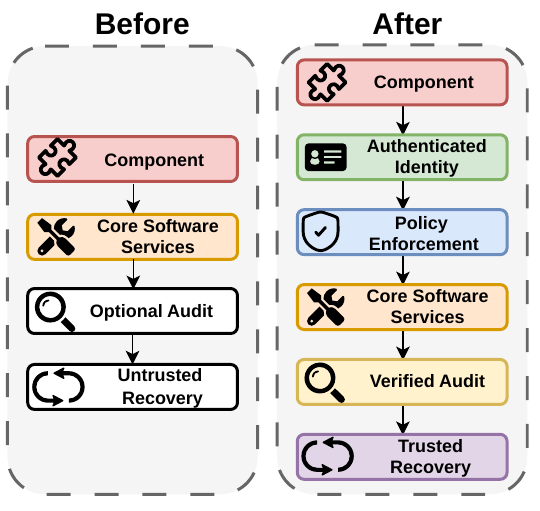}
    \caption{Recommended security architecture in which authenticated identity and policy enforcement isolate components from core software services, while verified auditing and trusted recovery detect misuse and restore the system when necessary.}
    \label{fig:security_rec}
\end{figure}

\subsection{Architectural Remediation}

\textbf{Authenticated identity and policy enforcement.}
Existing mechanisms protect communication links and message provenance~\cite{esiner2023f2pro}, but do not consistently bind internal operations to the application that initiated them. Building on these mechanisms, flight-software frameworks should preserve authenticated application identity across messages, services, hardware access, and persistent-state changes, then enforce authorization policies based on that identity.

\textbf{Isolation and bounded authority.}
Prior work demonstrates OS-level partitioning and capability-based access control~\cite{windsor2011partitioning,FurgalaPortingNC,UCAM-CL-TR-963}, providing mechanisms that can form the basis for stronger component isolation. Our findings indicate that isolation must also constrain application-specific authority over flight-software resources. Frameworks should therefore combine isolation with distinct credentials, resource restrictions, narrow interfaces, and capabilities limited to each component's required messages, services, devices, and persistent objects.

\textbf{Tamper-resistant accountability.}
Prior work supports protected logging and telemetry-based anomaly detection~\cite{bailey2021spacecraft,aerospace2024dars,ferretti2021auditable}, but these mechanisms do not necessarily provide mandatory attribution for security-sensitive operations across the flight-software architecture. Frameworks should extend these approaches with trusted audit points at framework, scheduler, communication, and hardware-access boundaries that record the authenticated requester, resource, action, and outcome in tamper-resistant logs.

\textbf{Trusted recovery and protected state.}
Prior work demonstrates secure boot, authenticated updates, rollback, and data recovery~\cite{bailey2021spacecraft,sparta2022secureboot,li2025aegissat,yang2022plc}, establishing important mechanisms for restoring software integrity. Our findings show that recovery must also account for security-relevant persistent state controlled by compromised components. Frameworks should therefore recover from an independently trusted baseline that verifies both security-relevant software and persistent state before normal operation resumes.

Taken together, these recommendations do not replace existing flight-software security mechanisms; they show how prior approaches can be extended and integrated to address the architectural weaknesses exposed by our analysis. Existing work provides many of the necessary building blocks, but those mechanisms must operate as part of a coherent architecture that constrains component authority across system boundaries. Development, testing, and certification remain necessary, but they should not be the only safeguards against a component exceeding its intended role. The flight-software architecture should itself constrain unauthorized actions, preserve reliable evidence of violations, protect security-relevant state, and provide a trusted path to recovery.

\subsection{Limitations}

\textbf{Construct Validity.} This work evaluates cFS more deeply than the other frameworks. The cross-framework comparison relies on published documentation, API references, maintained repositories, and targeted source inspection rather than equivalent experiments or comprehensive source audits. Consequently, differences in documentation quality and analysis depth may affect the consistency of the comparison. Undocumented mission-specific controls may also provide stronger boundaries than those identified here.

\textbf{External Validity.} The experiments were conducted in NOS3, which preserves the cFS application model, Software Bus interactions, and relevant APIs but does not reproduce every processor, timing, hardware-protection, communication, or recovery property of an operational spacecraft. The threat model further assumes that malicious code has already entered an integrated component. The results therefore characterize post-compromise capabilities and containment within the evaluated environment, rather than the likelihood or mechanism of initial compromise.

%% file: sections/conclusion.tex
\section{Conclusion} \label{sec:conclude}

Our results suggest that many previously identified flight software weaknesses are manifestations of a common underlying property: architectures establish trust implicitly and provide few mechanisms to enforce, constrain, or revoke it once components begin executing. By modeling trust across execution, identity, communication, observability, and persistence, we provide a systematic framework for reasoning about these behaviors and their security consequences.
Meaningful security cannot be achieved through incremental hardening alone. Security-critical guarantees should be enforced by the architecture itself rather than delegated to individual implementations or added later through mission-specific patches.
As satellites continue to support critical infrastructure for communication, navigation, and defense, establishing secure-by-construction flight software architectures is no longer optional. We hope that this work motivates a broader reevaluation of trust, governance, and security in next-generation space systems.

\clearpage